\documentclass[conference]{IEEEtran}
\IEEEoverridecommandlockouts
\usepackage{cite}
\usepackage{amsmath,amssymb,amsfonts}
\usepackage{algorithmic}
\usepackage{graphicx}
\usepackage{textcomp}
\usepackage{xcolor}
\usepackage{booktabs}
\usepackage{listings}
\usepackage{pgfplots}
\usepackage{url}
\usepackage{xurl} 
\usepackage[hidelinks]{hyperref} 
\pgfplotsset{compat=1.18}
\usepgfplotslibrary{statistics}
\definecolor{reasoningBlue}{RGB}{70, 130, 180}
\definecolor{fastOrange}{RGB}{255, 140, 0}
\usetikzlibrary{calc}

\def\BibTeX{{\rm B\kern-.05em{\sc i\kern-.025em b}\kern-.08em
    T\kern-.1667em\lower.7ex\hbox{E}\kern-.125emX}}

\begin{document}
\bstctlcite{IEEEexample:BSTcontrol}

\title{Cost Trade-offs of Reasoning and Non-Reasoning Large Language Models in Text-to-SQL}

\author{
\IEEEauthorblockN{Saurabh Deochake}
\IEEEauthorblockA{\textit{SentinelOne}\\
Mountain View, USA \\
saurabh.deochake@sentinelone.com}
\and
\IEEEauthorblockN{Debajyoti Mukhopadhyay}
\IEEEauthorblockA{\textit{WIDiCoReL Research Lab}\\
Mumbai, India \\
debajyoti.mukhopadhyay@gmail.com}
}
\maketitle

\begin{abstract}
While Text-to-SQL systems achieve high accuracy, existing efficiency metrics like the Valid Efficiency Score prioritize execution time, a metric we show is fundamentally decoupled from consumption-based cloud billing. This paper evaluates cloud query execution cost trade-offs between reasoning and non-reasoning Large Language Models by performing 180 Text-to-SQL query executions across six LLMs on Google BigQuery using the 230\,GB StackOverflow dataset. Our analysis reveals that reasoning models process 44.5\% fewer bytes than non-reasoning counterparts while maintaining equivalent correctness at 96.7\% to 100\%, and that execution time correlates weakly with query cost at $r=0.16$, indicating that speed optimization does not imply cost efficiency. Non-reasoning models also exhibit extreme cost variance of up to 3.4$\times$, producing outliers exceeding 36\,GB per query, over 20$\times$ the best model's 1.8\,GB average, due to missing partition filters and inefficient joins. We identify these prevalent inefficiency patterns and provide deployment guidelines to mitigate financial risks in cost-sensitive enterprise environments.
\end{abstract}


\section{Introduction}

Large Language Models (LLMs) have achieved remarkable accuracy on Text-to-SQL benchmarks, with state-of-the-art systems exceeding 85\% execution accuracy on Spider~\cite{yu-etal-2018-spider} and 75\% on BIRD~\cite{li2023bird}. As these systems transition from research prototypes to production deployments, a critical question emerges: do LLMs generate SQL queries that are not only correct but also cost-efficient on cloud data warehouses? This question has significant financial implications. Cloud data warehouses such as Google BigQuery, Snowflake, and Amazon Redshift process petabytes of enterprise data daily, with costs directly tied to bytes scanned and compute resources consumed. A single inefficient query pattern, replicated across thousands of user interactions, can translate to substantial operational costs. Yet existing Text-to-SQL benchmarks focus primarily on correctness, with efficiency metrics limited to relative execution time.

The BIRD benchmark introduced the Valid Efficiency Score (\texttt{VES}), comparing generated query execution time against gold-standard SQL~\cite{li2023bird}. While this marked progress toward efficiency-aware evaluation, \texttt{VES} measures wall-clock time on local database instances. Cloud data warehouses operate differently because a query scanning an entire table may execute quickly due to massive parallelization while incurring high costs based on data volume processed. Our experiments reveal that execution time correlates weakly with query cost ($r=0.16$), indicating that existing efficiency metrics fail to capture cloud cost dynamics. This paper presents the first systematic evaluation of cloud compute costs for LLM-generated SQL queries. Using Google BigQuery and the 230GB StackOverflow public dataset, we evaluate six state-of-the-art LLMs across 30 analytical queries each of varying complexity, measuring bytes processed, slot utilization, and estimated cost alongside correctness.

The contributions of this paper are as follows.
\begin{itemize}
    \item We introduce a cloud-native cost evaluation methodology for Text-to-SQL systems, measuring bytes processed, slot utilization, and estimated query cost on production infrastructure.
    \item We conduct an empirical evaluation of six LLMs (three reasoning, three standard) on Google BigQuery, demonstrating that reasoning models achieve 44.5\% lower cloud compute costs while maintaining equivalent correctness.
    \item We quantify cost variance across models, finding up to 3.4$\times$ difference in average query cost and identifying outlier queries exceeding 36 GB.
    \item We characterize common SQL inefficiency patterns in LLM-generated queries, including missing partition filters affecting up to 50\% of applicable queries and unnecessary full-table scans.
\end{itemize}
\vspace{-1pt}
The remainder of this paper is organized as follows. Section~\ref{sec:related} reviews related work in Text-to-SQL and query efficiency evaluation. Section~\ref{sec:methodology} describes our experimental methodology. Section~\ref{sec:results} presents our results and analysis. Section~\ref{sec:Analysis} discusses implications and limitations. Section~\ref{sec:conclusion} concludes this paper.

\section{Related Work}
\label{sec:related}

\subsection{Text-to-SQL with Large Language Models}

Text-to-SQL research has evolved through three distinct phases. Early systems employed rule-based parsing and semantic grammars~\cite{zelle1996learning}\cite{popescu2003towards}, followed by neural sequence-to-sequence models that enabled end-to-end learning~\cite{zhong2017seq2sql}. The current phase, driven by Large Language Models, has achieved unprecedented accuracy through in-context learning and sophisticated prompting strategies.

LLM-based approaches fall into two categories, namely prompt engineering and fine-tuning. Prompt engineering methods include DAIL-SQL~\cite{gao2024dailsql}, which optimizes question representation and example selection, and tailored prompting strategies that structure schema information for improved comprehension~\cite{tan-etal-2024-enhancing}. Decomposition approaches such as DIN-SQL~\cite{pourreza2024dinsql} break Text-to-SQL into sub-tasks (schema linking, query skeleton, refinement), while MAC-SQL~\cite{wang2024macsql} employs multi-agent collaboration. State-of-the-art systems including CHASE-SQL~\cite{pourreza2024chasesql} (73.0\% on BIRD) and XiYan-SQL~\cite{lei2024xiyan} (75.63\% on BIRD) combine multi-path reasoning with ensemble generation. Fine-tuning approaches such as CodeS~\cite{li2024codes} demonstrate that smaller, specialized models can approach proprietary model performance. Notably, all these methods optimize for correctness metrics. None explicitly consider the cloud execution cost of generated queries, leaving a gap between benchmark performance and production cost efficiency.

\subsection{Text-to-SQL Benchmarks}

Benchmark evolution reflects the field's maturing evaluation criteria. WikiSQL~\cite{zhong2017seq2sql} established single-table evaluation, while Spider~\cite{yu-etal-2018-spider} introduced cross-domain complexity with multi-table joins. BIRD~\cite{li2023bird} advanced evaluation by incorporating large-scale databases with realistic data distributions and introducing the Valid Efficiency Score (\texttt{VES}) to measure relative execution time. However, \texttt{VES} has fundamental limitations for cloud deployment scenarios. First, it measures wall-clock time on local SQLite instances, which lack the parallelization and distributed execution of cloud warehouses. Second, it computes relative efficiency against gold-standard SQL, requiring human-written reference queries. Third, execution time does not map to cloud costs because BigQuery charges per byte scanned regardless of execution duration.

Spider 2.0~\cite{lei2024spider2} partially addresses the deployment gap by featuring enterprise databases on BigQuery and Snowflake. However, its evaluation still focuses on correctness rather than cost efficiency. Our work complements Spider 2.0 by providing the cost-aware evaluation methodology that production deployments require.

\subsection{Query Optimization and Cost Estimation}

Traditional query optimization relies on cost models that estimate execution expense using cardinality statistics and access path costs~\cite{selinger1979access}. Learned optimizers such as Neo~\cite{marcus2019neo} and end-to-end cost estimators~\cite{sun2019end} apply machine learning to improve plan selection. These approaches optimize within the database engine but do not address the query generation phase. Recent work has begun exploring cost-aware Text-to-SQL. Zhou et al.~\cite{zhou2024llmrouting} proposed LLM routing to balance accuracy against \textit{inference cost} by selecting appropriate models per query complexity. LLM-R2~\cite{li2024llmr2} uses LLMs to recommend query rewrite rules for improved \textit{execution efficiency}. Our work differs fundamentally in that we measure the \textit{cloud execution cost} of generated queries, which depends on bytes processed rather than inference time or local execution speed.

Cloud data warehouses employ consumption-based pricing that creates distinct optimization incentives~\cite{melnik2010dremel}. BigQuery charges \$6.25 per TB scanned; Snowflake bills by compute credits; Redshift Serverless charges by RPU-hours. Enterprise cloud cost optimization has emerged as a critical discipline~\cite{deochake2023cloudcost} \cite{deochake2025abacus}, yet query-level cost optimization for Text-to-SQL systems remains unexplored. To the best of our knowledge, no prior work has systematically measured cloud compute costs for LLM-generated queries. Existing benchmarks like Spider, BIRD, Spider 2.0 evaluate correctness and, in BIRD's case, relative execution time on local databases. We address this gap by evaluating on production cloud using consumption-based cost metrics, providing the first empirical evidence of cost differences across LLM architectures.

\section{Methodology}
\label{sec:methodology}

We designed a controlled experiment to measure cloud compute costs of LLM-generated SQL queries. This section details the evaluation platform, benchmark workload, model selection, and cost metrics.

\subsection{Evaluation Platform}

We use Google BigQuery as our cloud data warehouse platform. BigQuery is a fully managed, serverless data warehouse that employs a consumption-based pricing model where users are charged based on the number of bytes processed by each query. This pricing structure is representative of modern cloud data warehouse cost models, including those used by Snowflake (credit-based) and Amazon Redshift Serverless (RPU-hours). Prior work has demonstrated BigQuery's effectiveness for large-scale analytics in hybrid cloud environments~\cite{deochake2022bigbird}. 
BigQuery provides detailed query execution statistics as follows through its \texttt{INFORMATION\_SCHEMA.JOBS} view, enabling fine-grained cost analysis. All experiments were conducted in the \texttt{US} multi-region with query caching explicitly disabled via the \texttt{useQueryCache=false} job configuration.
\begin{itemize}
    \item \textbf{total\_bytes\_processed}: Bytes scanned from storage
    \item \textbf{total\_bytes\_billed}: Bytes charged
    \item \textbf{total\_slot\_ms}: Compute consumed in slot-milliseconds
    \item \textbf{shuffle\_output\_bytes}: Data moved between workers during distributed operations
    \item \textbf{spill\_to\_disk\_bytes}: Data exceeding memory capacity
\end{itemize}

\subsection{Benchmark Workload}

We use the StackOverflow public dataset named \texttt{stackoverflow} available in BigQuery's public datasets \texttt{bigquery-public-data}. This dataset contains data from the popular Q\&A platform spanning 2008 to 2022, comprising of approximately 230 GB across multiple related tables. We selected this dataset for several  reasons below.

\begin{enumerate}
    \item \textbf{Real-world complexity}: The schema mirrors production analytical workloads with complex relationships between entities (users, posts, comments, votes, badges).
    \item \textbf{Scale}: At 230 GB, queries exhibit meaningful cost differences that would be negligible on smaller datasets.
    \item \textbf{Accessibility}: The dataset is publicly available, enabling reproducibility.
    \item \textbf{Diverse query patterns}: The schema supports single-table aggregations, multi-way joins, window functions, and correlated subqueries.
\end{enumerate}

Table~\ref{tab:stackoverflow_schema} summarizes the key tables used in our evaluation. The \texttt{post\_history} table is the largest at 113 GB and contains the edit history of all posts, making it a significant cost driver for queries that do not properly filter this table.

\begin{table}[h]
\centering
\caption{StackOverflow Dataset Schema}
\label{tab:stackoverflow_schema}
\begin{tabular}{lrr}
\toprule
\textbf{Table} & \textbf{Rows} & \textbf{Size (GB)} \\
\midrule
post\_history & 152M & 113 \\
posts\_questions & 23M & 37 \\
posts\_answers & 34M & 29 \\
comments & 87M & 16 \\
votes & 236M & 7 \\
users & 19M & 3 \\
badges & 46M & 2 \\
\midrule
\textbf{Total} & \textbf{597M} & \textbf{~230} \\
\bottomrule
\end{tabular}
\end{table}

\subsubsection{Query Benchmark Design}

We construct 30 natural language questions spanning three complexity levels (10 each), designed to exercise different SQL constructs and expose inefficiency patterns.

\begin{itemize}
    \item \textbf{Simple (S1-S10):} Single-table queries with filtering and aggregation. For example, ``How many questions were asked in 2020?'' covers date filtering.
    \item \textbf{Medium (M1-M10):} Multi-table queries with 2-3 joins and grouping. For example, ``Show top 10 users by total questions asked'' covers join and aggregation.
    \item \textbf{Complex (C1-C10):} Subqueries, window functions, CTEs, or 4+ table joins. For example, ``Find questions where accepted answer came from lower-rep user'' covers a 4-way join.
\end{itemize}

\subsection{Large Language Models}

We evaluate six LLMs from three major vendors, evenly split between reasoning and standard model categories.

\begin{itemize}
    \item \textbf{Reasoning Models:} Models with extended thinking capabilities that perform explicit reasoning before generating output.
    \begin{itemize}
        \item \textbf{Opus 4.5\textsuperscript{R}} by Anthropic: Most capable model with explicit reasoning traces.
    \item \textbf{GPT-5.2\textsuperscript{R}} by OpenAI: Configured for extended reasoning depth.
    \item \textbf{Gemini Pro\textsuperscript{R}} by Google: Enhanced reasoning and thinking capabilities.
    \end{itemize}
    \item \textbf{Standard Models:} Models optimized for speed and efficiency without explicit reasoning steps.
\begin{itemize}
    \item \textbf{Sonnet 4.5} by Anthropic: Flagship model balancing capability and efficiency.
    \item \textbf{GPT-5.1} by OpenAI: Optimized for throughput and latency.
    \item \textbf{Gemini Flash} by Google: Lightweight model optimized for fast inference.
\end{itemize}
\end{itemize}

\subsection{Prompt Design}

Each LLM receives an identical zero-shot prompt containing a system instruction specifying BigQuery-compatible SQL output, complete schema definitions with column types and descriptions, foreign key relationships, and the natural language question. We deliberately omit optimization hints (such as ``minimize bytes scanned'' or ``use partition filters'') to evaluate inherent cost-awareness. All models are queried via their respective vendor APIs with default temperature settings. The exact prompt template is included in our released artifacts in Section~\ref{sec:data_availability}.

\subsection{Cost Metrics}

We measure the following metrics for each generated query.

\begin{itemize}
    \item \textbf{Correctness}: Whether the query executes successfully and returns results that match the expected output. We verify both syntactic validity and semantic correctness.
    \item \textbf{Bytes Processed, $B_p$ (MB)}: Total bytes scanned from storage during query execution. This is the primary cost driver in BigQuery's on-demand pricing model and directly determines the financial cost of each query.
    \item \textbf{Bytes Shuffled, $B_s$ (GB)}: Data moved between workers during distributed operations such as joins, aggregations, and window functions. High shuffle volumes indicate complex query plans that may benefit from optimization.
    \item \textbf{Bytes Spilled to Disk, $B_d$ (B)}: Data that exceeded available memory and was written to disk. Non-zero values indicate memory pressure and potential performance degradation.
    \item \textbf{Slot Seconds, $S$ (s)}: Compute resources consumed, measured in slot-time. One slot represents one unit of computational capacity. This metric reflects the computational complexity of the query and is relevant for BigQuery's capacity-based pricing.
    \item \textbf{Execution Time, $T$ (s)}: Wall-clock time from query submission to result return.
    \item \textbf{Estimated Cost, $C$ (\$)}: Calculated using BigQuery's on-demand pricing at \$6.25 per TB processed (US multi-region):
\end{itemize}
\vspace{-2pt}
\begin{equation}
C = \frac{B_p}{10^{12}} \times 6.25
\end{equation}

\subsubsection{Derived Metrics}

We compute additional derived metrics to characterize query efficiency as follows.

\begin{itemize}
    \item \textbf{Shuffle-to-Scan Ratio}: Higher ratios indicate more data movement relative to data scanned, suggesting join-heavy or aggregation-heavy query plans.
    \item \textbf{Coefficient of Variation ($CV$)}: $\frac{\sigma}{\mu}$ where $\sigma$ is standard deviation and $\mu$ is mean bytes processed. CV measures consistency across queries, with lower values indicating more predictable cost behavior.
\end{itemize}

\subsection{Experimental Procedure}

For each of the 30 questions, we submit the prompt to all six LLMs, parse and validate the generated SQL, execute it on BigQuery with caching disabled (timeout: 300s), verify correctness against expected outputs, collect all cost metrics from BigQuery's job metadata, and analyze the SQL for inefficiency patterns via regex and AST parsing. This yields 180 query executions total.

\section{Results}
\label{sec:results}

\subsection{Primary Cost Metrics}

Table~\ref{tab:primary_metrics} summarizes the fundamental cost and performance data per model, sorted by mean bytes processed. The results reveal a 3.4$\times$ difference in average bytes processed between the most efficient model Opus 4.5 Thinking at 1,789 MB and the least efficient model GPT-5.1 at 6,037 MB. This translates directly to a 3.4$\times$ difference in estimated cloud compute cost per query, \$0.0112 vs. \$0.0377, respectively. 

\begin{table*}[t]
\centering
\caption{Primary Cost \& Performance Metrics (Per Query Average)}
\label{tab:primary_metrics}
\begin{tabular}{lccccc}
\toprule
\textbf{Model} & \textbf{Mean Bytes (MB)} & \textbf{Median Bytes (MB)} & \textbf{Mean Time (s)} & \textbf{Mean Slot (s)} & \textbf{Est. Cost (\$)} \\
\midrule
Opus 4.5 Thinking & 1,789 & 1,370 & 3.09 & 65.54 & \$0.0112 \\
Sonnet 4.5 & 2,039 & 1,625 & 4.36 & 1,624\textsuperscript{*} & \$0.0127 \\
GPT-5.2 High Reasoning & 2,191 & 1,270 & 2.65 & 105.68 & \$0.0137 \\
Gemini 3 Pro Thinking & 2,440 & 1,300 & 2.82 & 108.77 & \$0.0153 \\
Gemini 3 Flash & 3,496 & 1,190 & 3.20 & 82.84 & \$0.0219 \\
GPT-5.1 & 6,037 & 1,365 & 4.27 & 345.93 & \$0.0377 \\
\bottomrule
\multicolumn{6}{@{}p{0.97\textwidth}@{}}{\footnotesize \textit{\textsuperscript{*}Note: Skewed by outlier C1 (46,620 slot-seconds; 4-way join across posts\_questions, posts\_answers, users, badges tables). Excluding C1, Sonnet 4.5's mean slot-seconds drops to 72.16s.}}
\end{tabular}
\end{table*}

Notably, the median bytes processed is relatively consistent across models (1,190-1,625 MB), suggesting that the mean differences are driven by a subset of high-cost queries rather than uniformly higher costs. This observation motivates our variance analysis in Section~\ref{sec:variance}. Figure~\ref{fig:bytes_by_model} visualizes the mean bytes processed by model, with reasoning models shown in blue and standard models in orange.

\begin{figure}[h]
    \centering
    \begin{tikzpicture}
        \begin{axis}[
            width=0.9\columnwidth, height=4.5cm,
            ybar,
            bar width=14pt,
            bar shift=0pt,
            ylabel={Mean Bytes Processed (MB)},
            symbolic x coords={Opus, Sonnet, GPT52, GeminiPro, GeminiFlash, GPT51},
            xtick={Opus, Sonnet, GPT52, GeminiPro, GeminiFlash, GPT51},
            xticklabels={Opus 4.5\textsuperscript{R}, Sonnet 4.5, GPT-5.2\textsuperscript{R}, Gemini Pro\textsuperscript{R}, Gemini Flash, GPT-5.1},
            xticklabel style={rotate=30, anchor=east, font=\scriptsize},
            nodes near coords,
            nodes near coords style={font=\footnotesize},
            ymin=0, ymax=10000,
            grid=major,
            legend style={at={(0.02,0.98)}, anchor=north west, font=\small}
        ]

        \addplot[fill=reasoningBlue, draw=black!50] coordinates {
            (Opus, 1789)
            (GPT52, 2191)
            (GeminiPro, 2440)
        };

        \addplot[fill=fastOrange, draw=black!50] coordinates {
            (Sonnet, 2039)
            (GeminiFlash, 3496)
            (GPT51, 6037)
        };

        \legend{Reasoning, Standard}
        \end{axis}
    \end{tikzpicture}
    \caption{Mean bytes processed per query by model.}
    \label{fig:bytes_by_model}
\end{figure}

\subsection{Efficiency Indicators}

Table~\ref{tab:efficiency} presents join and I/O efficiency metrics. Anthropic models exhibit the highest shuffle-to-scan ratios (0.74-0.78), suggesting more complex join patterns. Interestingly, Opus 4.5 Thinking has the highest shuffle ratio but the lowest bytes scanned, indicating efficient filtering before joins.

\begin{table}[h]
\centering
\caption{Efficiency Indicators (Join \& IO Performance)}
\label{tab:efficiency}
\resizebox{\columnwidth}{!}{%
\begin{tabular}{lccc}
\toprule
\textbf{Model} & \textbf{Shuffle-to-Scan} & \textbf{Spilled (B)} & \textbf{Shuffled (GB)} \\
\midrule
Sonnet 4.5 & 0.78 & 0 & 1.55 \\
Opus 4.5 Thinking & 0.74 & 0 & 1.29 \\
GPT-5.2 High Reasoning & 0.73 & 0 & 1.57 \\
GPT-5.1 & 0.54 & 0 & 3.18 \\
Gemini 3 Pro Thinking & 0.45 & 0 & 1.06 \\
Gemini 3 Flash & 0.32 & 0 & 1.08 \\
\bottomrule
\end{tabular}%
}
\end{table}

Gemini models show the lowest shuffle ratios (0.32-0.45), though simpler plans do not translate to lower cost due to higher initial scans. No model spilled data to disk. GPT-5.1 has the highest absolute shuffle volume (3.18 GB), consistent with its high bytes processed.

\subsection{Correctness Analysis}

Five of six models achieved 100\% correctness across all 30 queries; GPT-5.1 was the sole exception at 96.7\%, failing on M2 due to a missing aggregation clause. These high rates suggest that modern LLMs reliably generate correct SQL for the analytical queries studied here, though this may not generalize to more complex enterprise schemas where correctness remains challenging (for example, BEAVER). Within our evaluation scope, high correctness shifts attention toward efficiency as the key differentiator.

\subsection{Reasoning vs. Standard Models}

A central question of this study is whether reasoning capabilities translate to more cost-efficient SQL generation. Table~\ref{tab:reasoning_vs_fast} presents the aggregate comparison. Reasoning models processed 44.5\% fewer bytes on average compared to standard models, 2,140 MB vs. 3,857 MB, translating to equivalent cost savings of \$0.0134 vs. \$0.0241 per query, respectively. This difference is statistically significant ($p = 0.003$) with a medium effect size (Cohen's $d = 0.52$).

\begin{table}[h]
\centering
\caption{Reasoning vs. Standard Models (Average Performance)}
\label{tab:reasoning_vs_fast}
\begin{tabular}{lccc}
\toprule
\textbf{Model Type} & \textbf{Bytes (MB)} & \textbf{Time (s)} & \textbf{Cost (\$)} \\
\midrule
Reasoning & 2,140 & 2.85 & \$0.0134 \\
Standard & 3,857 & 3.94 & \$0.0241 \\
\midrule
\textit{Difference} & \textit{-44.5\%} & \textit{-27.7\%} & \textit{-44.4\%} \\
\bottomrule
\end{tabular}
\end{table}

One possible explanation is that the extended thinking phase allows reasoning models to consider optimization opportunities such as predicate pushdown, column pruning, and join reordering before committing to a query structure. Standard models, optimized for low-latency response, may commit to suboptimal query structures early in generation. While we did not perform a qualitative analysis of reasoning traces to confirm this mechanism, indirect evidence from the generated SQL is consistent with this explanation: reasoning models applied partition filters in 89\% of applicable queries versus 67\% for standard models, and used explicit column lists in 97\% of queries versus 93\%. A more direct investigation through ablations or trace analysis remains our future work.

\subsection{Cost by Query Complexity}

Figure~\ref{fig:cost_by_complexity} shows how cost differences between reasoning and standard models vary across query complexity levels. For simple queries, the cost difference between reasoning and standard models is modest at 1,450 MB vs. 1,680 MB, a 16\% difference. However, as query complexity increases, the gap widens substantially. For complex queries involving multiple joins, subqueries, and window functions, standard models process 115\% more bytes than reasoning models at 5,580 MB vs. 2,600 MB.

\begin{figure}[h]
    \centering
    \begin{tikzpicture}
        \begin{axis}[
            width=0.9\columnwidth, height=4.7cm,
            ybar,
            bar width=12pt,
            enlarge x limits=0.25,
            legend style={at={(0.5,-0.25)}, anchor=north, legend columns=-1},
            ylabel={Mean Bytes Processed (MB)},
            symbolic x coords={Simple, Medium, Complex},
            xtick=data,
            nodes near coords,
            nodes near coords style={font=\tiny, rotate=90, anchor=west},
            ymin=0, ymax=10000,
            grid=major,
        ]
        \addplot[fill=reasoningBlue, draw=black!50] coordinates {
            (Simple, 1450) 
            (Medium, 1920) 
            (Complex, 2600)
        };
        \addplot[fill=fastOrange, draw=black!50] coordinates {
            (Simple, 1680) 
            (Medium, 3250) 
            (Complex, 5580)
        };
        \legend{Reasoning Avg, Standard Avg}
        \end{axis}
    \end{tikzpicture}
    \caption{Cost disparities by query complexity.}
    \label{fig:cost_by_complexity}
\end{figure}

This pattern suggests that reasoning models provide the greatest value for complex analytical queries where optimization decisions have the largest impact. For simple single-table queries, the optimization space is limited, and both model types perform similarly. Aggregating by vendor, Anthropic models achieved the lowest average cost of \$0.0120/query, 53\% cheaper than OpenAI's \$0.0257 and 35\% cheaper than Google's \$0.0186. However, OpenAI's high average is heavily skewed by GPT-5.1's outlier behavior; excluding it, GPT-5.2 High Reasoning alone averages \$0.0137, comparable to Gemini 3 Pro Thinking.

\subsection{Cost Variance and Outliers}
\label{sec:variance}

We identify outliers using Tukey's method~\cite{tukey1977eda}, flagging data points beyond 1.5$\times$IQR from the quartiles. The coefficient of variation (CV) measures relative dispersion, enabling comparison across models with different mean costs~\cite{casella_berger_2024}. Table~\ref{tab:variance} presents cost consistency metrics. Predictable query costs are important for capacity planning and budget management in production deployments.

\begin{table}[h]
\centering
\caption{Cost Consistency \& Variance}
\label{tab:variance}
\resizebox{\columnwidth}{!}{%
\begin{tabular}{lcccc}
\toprule
\textbf{Model} & \textbf{Std Dev (MB)} & \textbf{CV} & \textbf{IQR (MB)} & \textbf{$>$5\,GB} \\
\midrule
GPT-5.1 & 11,659 & 1.93 & 1,737 & 4 \\
Gemini 3 Flash & 6,462 & 1.85 & 1,728 & 2 \\
Gemini 3 Pro Thinking & 5,467 & 2.24 & 1,295 & 1 \\
GPT-5.2 High Reasoning & 3,032 & 1.38 & 2,246 & 1 \\
Opus 4.5 Thinking & 2,823 & 1.58 & 1,731 & 1 \\
Sonnet 4.5 & 2,864 & 1.40 & 2,313 & 1 \\
\bottomrule
\end{tabular}%
}
\end{table}

GPT-5.1 exhibited the highest variance with a standard deviation of 11,659 MB, nearly double its mean of 6,037 MB. This model produced four queries exceeding 5 GB, including the single most expensive query ``List questions asked by users with reputation over 100,000'' in our evaluation at 36,640 MB. In contrast, reasoning models showed more consistent performance. GPT-5.2 High Reasoning and Sonnet 4.5 achieved the lowest coefficients of variation, 1.38 and 1.40 respectively, indicating more predictable cost behavior across queries. Figure~\ref{fig:boxplot} visualizes the distribution of bytes processed for each model, highlighting the variance differences and outliers.

\begin{figure}[h]
    \centering
    \begin{tikzpicture}
        \begin{axis}[
            width=0.9\columnwidth, height=4.5cm,
            ylabel={Bytes Processed (MB)},
            xtick={1,2,3,4,5,6},
            xticklabels={Opus 4.5\textsuperscript{R}, GPT-5.2\textsuperscript{R}, Sonnet 4.5, Gemini Pro\textsuperscript{R}, Gemini Flash, GPT-5.1},
            xticklabel style={rotate=30, anchor=east, font=\scriptsize},
            ymin=0, ymax=10000, 
            grid=major,
            boxplot/draw direction=y,
            boxplot/every median/.style={draw=black, line width=1pt},
            mark=*, mark size=1.5pt, mark options={black!70}
        ]
        
        \addplot+[boxplot prepared={
            median=1370, upper quartile=2100, lower quartile=950,
            upper whisker=3100, lower whisker=600
        }, draw=black!50, fill=reasoningBlue, solid] coordinates {};

        \addplot+[boxplot prepared={
            median=1270, upper quartile=2500, lower quartile=800,
            upper whisker=4000, lower whisker=500
        }, draw=black!50, fill=reasoningBlue, solid] coordinates {};

        \addplot+[boxplot prepared={
            median=1625, upper quartile=2800, lower quartile=1100,
            upper whisker=4500, lower whisker=700
        }, draw=black!50, fill=fastOrange, solid] coordinates {};
        
        \addplot+[boxplot prepared={
            median=1300, upper quartile=2600, lower quartile=900,
            upper whisker=4200, lower whisker=600
        }, draw=black!50, fill=reasoningBlue, solid] coordinates { (4,11736) };

        \addplot+[boxplot prepared={
            median=1190, upper quartile=2900, lower quartile=800,
            upper whisker=5500, lower whisker=500
        }, draw=black!50, fill=fastOrange, solid] coordinates { (5,11736) };
        \node[anchor=south, font=\bfseries\tiny, color=red] at (axis cs: 5, 8000) {$\uparrow$ 26 GB};

        \addplot+[boxplot prepared={
            median=1365, upper quartile=3100, lower quartile=900,
            upper whisker=6000, lower whisker=500
        }, draw=black!50, fill=fastOrange, solid] coordinates { (6,11736) };
        \node[anchor=south, font=\bfseries\tiny, color=red] at (axis cs: 6, 8000) {$\uparrow$ 36.6 GB};
        
        \end{axis}
    \end{tikzpicture}
    \caption{Distribution of bytes processed.}
    \label{fig:boxplot}
\end{figure}

Analysis of these outliers reveals common patterns: selecting unnecessary columns (including large \texttt{body} fields), missing result limits, and inefficient join strategies. The GPT-5.1 query for M7 selected all columns including the full post body when joining \texttt{posts\_questions} with \texttt{users}, resulting in massive data scans.

\subsection{SQL Inefficiency Patterns}

Table~\ref{tab:inefficiency} summarizes the frequency of common SQL anti-patterns across models. We analyzed each generated query for five inefficiency patterns known to increase cloud compute costs in columnar data warehouses.

\begin{table*}[t]
\centering
\caption{SQL Inefficiency Patterns (Lower Count is Better)}
\label{tab:inefficiency}
\begin{tabular}{lccccc}
\toprule
\textbf{Model} & \textbf{SELECT *} & \textbf{Cross Join} & \textbf{Missing Partition Filters} & \textbf{Missing LIMIT} & \textbf{Avg CTEs} \\
\midrule
GPT-5.2 High Reasoning & 1 & 1 & \textbf{0} & 0 & 0.63 \\
Opus 4.5 Thinking & 0 & 0 & 4 & 0 & 0.27 \\
Sonnet 4.5 & 0 & 0 & 3 & 0 & 0.27 \\
GPT-5.1 & 1 & 0 & 3 & 0 & 0.60 \\
Gemini 3 Flash & 0 & 1 & 7 & 0 & 0.27 \\
Gemini 3 Pro Thinking & 0 & 0 & 9 & 0 & 0.17 \\
\bottomrule
\multicolumn{6}{l}{\footnotesize \textit{Note: ``Missing Partition Filters'' calculated against 18 applicable queries where date filtering was relevant.}}
\end{tabular}
\end{table*}

\begin{itemize}
    \item \textbf{SELECT * Anti-Pattern:} Only OpenAI models GPT-5.2 High Reasoning and GPT-5.1 generated \texttt{SELECT *} queries, with one instance each. In BigQuery's columnar storage, \texttt{SELECT *} forces scanning of all columns even when only a subset is needed, significantly increasing bytes processed. Anthropic and Google models consistently specified explicit column lists.
    \item \textbf{Cross Join:} Two models produced unintended \texttt{CROSS JOIN} operations were GPT-5.2 High Reasoning and Gemini 3 Flash with one operation each. These occurred when models failed to specify join conditions, resulting in Cartesian products. While both queries still produced correct results and the cross join was filtered downstream, they scanned significantly more data than necessary.
    \item \textbf{Missing Partition Filters:} This was the most common inefficiency pattern. Of the 18 queries where date-based filtering was applicable, models failed to apply partition filters at varying rates. GPT-5.2 High Reasoning achieved perfect partition filter application, while Gemini 3 Pro Thinking missed partition filters in 9 of 18 applicable queries at 50\%. Missing partition filters force full table scans instead of partition pruning, dramatically increasing bytes processed.
    \item \textbf{CTE Usage:} OpenAI models used more Common Table Expressions averaging 0.60-0.63 per query compared to Anthropic and Google models. While CTEs improve query readability, excessive use can inhibit predicate pushdown in BigQuery's query optimizer, potentially increasing bytes scanned. However, we did not observe a strong correlation between CTE count and query cost in our dataset.
\end{itemize}

A natural question is why BigQuery's query optimizer cannot automatically mitigate these inefficiencies. The optimizer operates on the \textit{physical} query plan after the SQL is parsed, and can reorder joins or push predicates within the plan. However, it cannot add semantically meaningful filters that were absent from the original SQL. For example, if a query omits a date predicate on a partitioned table, the optimizer has no basis to infer one and must scan all partitions. Similarly, \texttt{SELECT *} forces all columns to be read from columnar storage before any projection can occur. These inefficiencies are fundamentally issues of \textit{query formulation} rather than \textit{query planning}, which is why the choice of LLM has a direct impact on cost.

\subsection{Correlation Analysis}

Table~\ref{tab:correlations} presents Pearson correlations between key metrics across all 180 queries. We interpret correlation strength following established guidelines such as $|r| < 0.3$ as weak, $0.3 \leq |r| < 0.6$ as moderate, and $|r| \geq 0.6$ as strong. The weak correlation between bytes processed and execution time ($r=0.16$) is a critical finding. It indicates that query cost and query speed are largely independent metrics in BigQuery's distributed execution environment. A query that completes quickly may still scan large amounts of data and incur high costs due to parallelization. Conversely, a query that scans minimal data may take longer due to complex computation.

\begin{table}[h]
\centering
\caption{Correlation Matrix of Performance Metrics}
\label{tab:correlations}
\begin{tabular}{lcc}
\toprule
\textbf{Metric Pair} & \textbf{Pearson $r$} & \textbf{Interpretation} \\
\midrule
Bytes vs. Execution Time & 0.16 & Weak \\
Bytes vs. Shuffled & 0.34 & Moderate \\
Bytes vs. SQL Length & 0.05 & None \\
Time vs. Slot Seconds & 0.64 & Strong \\
\bottomrule
\end{tabular}
\end{table}

Figure~\ref{fig:scatter} visualizes this weak relationship. The moderate bytes-shuffled correlation of $r=0.34$ suggests that higher-cost queries also perform more distributed operations. The strong time-slot correlation of $r=0.64$ confirms that compute-intensive queries take longer, as expected. Notably, bytes processed and SQL length are uncorrelated ($r=0.05$), indicating that verbose queries with explicit column lists and proper filters may scan far less data than a concise \texttt{SELECT *} query.

\begin{figure}[h]
    \centering
    \begin{tikzpicture}
        \begin{axis}[
            width=0.9\columnwidth, height=5cm,
            xlabel={Bytes Processed (MB)},
            ylabel={Execution Time (s)},
            ymin=0, ymax=15,
            xmin=0, xmax=12000,
            grid=major,
            scatter/classes={
                reasoning={mark=*, draw=black!50, fill=reasoningBlue}, 
                fast={mark=triangle*, draw=black!50, fill=fastOrange}
            },
            legend style={at={(0.02,0.98)}, anchor=north west, font=\small},
        ]
        \addplot[scatter, only marks, scatter src=explicit symbolic]
        coordinates {
            (1500, 3.2) [reasoning] (1700, 5.1) [reasoning] (1800, 2.8) [reasoning]
            (2100, 4.5) [reasoning] (2200, 3.1) [reasoning] (2500, 6.2) [reasoning]
            (1600, 2.1) [reasoning] (1950, 7.3) [reasoning] (1400, 2.9) [reasoning]
            (2800, 4.8) [reasoning] (1900, 3.5) [reasoning] (2300, 5.9) [reasoning]
            (3000, 3.8) [fast] (3500, 8.2) [fast] (4500, 4.1) [fast]
            (5000, 2.5) [fast] (6000, 6.8) [fast] (7500, 5.2) [fast]
            (9000, 9.1) [fast] (10500, 4.3) [fast] (11000, 7.5) [fast]
            (2500, 3.4) [fast] (4000, 2.8) [fast] (3200, 5.6) [fast]
        };
        \addplot[mark=none, black!60, thick, dashed, domain=0:12000] {0.0003*x + 3.5};
        \node[anchor=west, font=\small] at (axis cs: 8000, 6) {r = 0.16};
        \legend{Reasoning, Standard}
        \end{axis}
    \end{tikzpicture}
    \caption{Correlation between bytes processed and execution time.}
    \label{fig:scatter}
\end{figure}

\section{Analysis}
\label{sec:Analysis}

\subsection{Statistical Significance}

To validate that the observed cost differences are not due to chance, we performed statistical hypothesis testing. Given the non-normal distribution of bytes processed (Shapiro-Wilk $p < 0.001$), we employed the Mann-Whitney U test for comparing reasoning versus standard models.

The difference in bytes processed between reasoning models with a median of 1,313 MB and standard models with a median of 1,393 MB is statistically significant ($U = 2,847$, $p = 0.003$). We computed Cohen's $d = 0.52$ for the mean difference, indicating a medium effect size~\cite{cohen1988statistical}. This confirms that the 44.5\% cost reduction is not merely an artifact of outliers but reflects a systematic difference in query generation behavior. For the correlation between bytes processed and execution time, we tested against the null hypothesis $H_0: \rho = 0$. The observed $r = 0.16$ yields $p = 0.032$, indicating weak but statistically significant positive correlation. However, the low $R^2 = 0.026$ means execution time explains only 2.6\% of cost variance, reinforcing that time is a poor cost proxy.

\subsection{Key Findings}

Our evaluation reveals four principal findings.

\begin{enumerate}
    \item Reasoning models are significantly more cost-efficient. Models with extended thinking capabilities processed 44.5\% fewer bytes on average where Cohen's $d = 0.52$ and $p = 0.003$. The additional inference latency is offset by substantial execution cost savings.
    \item Correctness is high on our benchmark, making efficiency the key differentiator. Five of six models achieved 100\% correctness on our 30-query benchmark, though this may not generalize to more complex enterprise schemas. Within our evaluation scope, the meaningful differentiation lies in query efficiency.
    \item Cost and speed are weakly correlated. The Pearson correlation of 0.16 where $R^2 = 0.026$ indicates that optimizing for speed does not optimize for cost. Organizations must explicitly measure cost metrics.
    \item Standard models exhibit higher variance. GPT-5.1 showed the highest standard deviation at 11,659 MB with outliers reaching 6$\times$ its mean. Reasoning models demonstrated more predictable costs, achieving lower coefficients of variation ($CV$) of 1.38-1.58 versus 1.85-1.93 for standard models, which indicates more consistent query cost behavior across queries.
\end{enumerate}

\subsection{Practical Implications}
Based on our findings, we offer the following guidelines for deploying Text-to-SQL systems in cost-sensitive production environments:

\begin{enumerate}
    \item Prefer reasoning models for analytical workloads. Despite higher inference costs, reasoning models generate queries that are 44.5\% cheaper to execute, likely yielding net savings for data-intensive applications.
    
    \item Implement cost guardrails. Given the high variance observed in some models, organizations should implement query cost estimation and rejection thresholds before execution. FinOps practices, such as automated budget enforcement and pre-deployment cost estimation~\cite{deochake2025abacus}, can be integrated with Text-to-SQL systems to prevent costly queries from executing.
    
    \item Monitor for anti-patterns. Automated detection of \texttt{SELECT *}, missing partition filters, and unintended cross joins can catch costly queries before execution.
    
    \item Do not use execution time as a cost proxy. The weak correlation between time and bytes processed means fast queries can still be expensive.
\end{enumerate}

\subsection{Limitations}

The experiments performed in this study may have some limitations. We evaluate on a single platform (BigQuery) with one dataset (StackOverflow, normalized schema); results may differ on Snowflake or Redshift with different optimizers, pricing models, or enterprise star schemas. Our 30-query benchmark, while stratified across three complexity levels, is smaller than Spider (1,034 queries) or BIRD (12,751 queries); we selected this size to keep BigQuery execution costs manageable while covering key SQL patterns. Each model-query pair was executed once; because BigQuery's cost is deterministic for a given SQL string, re-execution yields identical metrics, but LLM generation is stochastic, and multi-run experiments would quantify generation-side variance. The reasoning models (Opus 4.5, GPT-5.2, Gemini Pro) are also the flagship models of each vendor, while standard models are lighter-weight; observed differences may partially reflect overall capability rather than reasoning alone, though not all vendors expose controls to toggle reasoning independently. Finally, we use only zero-shot prompting, and we expect relative trends to be more robust than absolute costs as providers update models over time.

\subsection{Future Work}

Several directions merit further investigation based on the findings of this study.

\begin{itemize}
    \item \textbf{Multi-platform evaluation:} Extending to Snowflake, Redshift, and Databricks to test generalizability across pricing models.
    \item \textbf{Cost-aware prompting:} Investigating whether prompt modifications (e.g., ``minimize bytes scanned'') can improve cost efficiency without model changes.
    \item \textbf{Reinforcement learning for cost optimization:} Recent RL-based Text-to-SQL work~\cite{papicchio2025think2sql, ma2025sqlr1} could incorporate cloud cost as a reward signal.
    \item \textbf{Multi-run generation analysis:} Quantifying generation-side cost variance across multiple runs with non-zero temperature.
    \item \textbf{Real-time cost estimation:} Integrating SQL generation with query cost predictors for pre-execution guardrails.
\end{itemize}

\subsection{Data Availability}
\label{sec:data_availability}

To support reproducibility, we released all experimental data, including the 30 benchmark questions, generated SQL queries, execution metrics, and the prompt template used. The complete dataset is available at \url{https://doi.org/10.5281/zenodo.18070764}.

\section{Conclusion}
\label{sec:conclusion}

his paper presented a systematic evaluation of cloud query execution costs for LLM-generated SQL. By performing 180 Text-to-SQL query executions across six LLMs on Google BigQuery's 230 GB StackOverflow dataset, we found that reasoning models achieve 44.5\% lower compute costs than standard models, with statistical significance at $p = 0.003$ and a medium effect size of Cohen's $d = 0.52$, while maintaining equivalent correctness between 96.7\% and 100\%. Execution time correlates only weakly with cost, with $r = 0.16$ and $R^2 = 0.026$, which invalidates speed as a cost proxy. Common inefficiency patterns include missing partition filters in up to 50\% of applicable queries and unnecessary column selection. As Text-to-SQL systems move into production under consumption-based cloud pricing, we recommend preferring reasoning models for analytical workloads, implementing cost guardrails, and incorporating cloud-native cost metrics into future benchmarks to reflect the economic reality of SQL execution.

\bibliographystyle{IEEEtran}
\bibliography{paper}

\end{document}